\documentclass[aps,pra,groupedaddress,amsmath,amssymb,twocolumn,nofootinbib,10pt,a4paper,showpacs]{revtex4-1}
\usepackage{hyperref}
\usepackage{graphicx,dsfont,amsthm}
\usepackage{color}

\usepackage{txfonts} 

\input{Qcircuit}

\newcommand{\ket}[1]{\left\vert{#1}\right\rangle}

\newcommand{\proj}[1]{\left\vert{#1}\right\rangle \left\langle{#1}\right\vert}

\newcommand{\trz}[2]{\mathrm{tr}_{#2}{\left(#1\right)}}

\newcommand{\braccket}[3]{{\langle #1 | #2 | #3 \rangle}}
\newcommand{\ketbra}[2]{{|#1\rangle\!\langle#2|}}
\newcommand{\id}{\mathds{1}}

\newcommand{\cnot}{\textsc{cnot}}
\newcommand{\cnota}[2]{\textsc{cnot}_{#1\rightarrow #2}}

\begin{document}

\newtheorem{thm}{Theorem}
\newtheorem{lem}{Lemma}
\newtheorem{defi}{Definition}

\title{Secret key rates for an encoded quantum repeater}
\author{Sylvia Bratzik}
\email{bratzik@thphy.uni-duesseldorf.de}
\author{Hermann Kampermann}
\author{Dagmar Bru{\ss}}
\affiliation{Institute for Theoretical Physics III, Heinrich-Heine-Universit\"at D\"usseldorf, 40225 D\"usseldorf, Germany.}

\date{\today}

\begin{abstract}
We investigate secret key rates for the quantum repeater using encoding [L.\ Jiang \textit{et al.}, Phys.\ Rev.\ A \textbf{79}, 032325 (2009)] and compare them to the standard repeater scheme by Briegel, D\"{u}r, Cirac, and Zoller. The former scheme has the advantage of a minimal consumption of classical communication. We analyze the trade-off in the secret key rate between the communication time and the required resources. For this purpose, we introduce an error model for the repeater using encoding which allows for input Bell states with a fidelity smaller than one, in contrast to the model given in [L.\ Jiang \textit{et al.}, Phys.\ Rev.\ A \textbf{79}, 032325 (2009)]. We show that one can correct additional errors in the encoded connection procedure of this repeater and develop a suitable decoding algorithm. Furthermore, we derive the rate of producing entangled pairs for the quantum repeater using encoding and give the minimal parameters (gate quality and initial fidelity) for establishing a nonzero secret key. We find that the generic quantum repeater is optimal regarding the secret key rate per memory per second and show that the encoded quantum repeater using the simple three-qubit repetition code can even have an advantage with respect to the resources compared to other recent quantum repeater schemes with encoding. 
\end{abstract}



\maketitle

\section{Introduction and motivation}

Quantum repeaters \cite{Briegel1998,Dur1999} are a necessary tool for long-distance quantum communication. They permit to establish entangled Bell pairs over distances of several hundreds of kilometers. A quantum repeater setup consists of segments of small distances, where entangled Bell pairs are created and then entanglement swapping \cite{Zukowski1993} with the neighboring pairs is performed. In order to overcome the decrease in the fidelity due to swapping, entanglement distillation \cite{Bennett1996a,Deutsch1996} can be performed. There are several suggestions for experimental realizations of quantum repeaters \cite{Duan2001,VanLoock2006}. For a recent review on quantum repeaters, see \cite{Sangouard2011}. 

Recently, we investigated the optimal quantum repeater setups with respect to the secret key rate \cite{Abruzzo2013,Bratzik2013}. The limiting factor of these quantum repeater schemes, especially regarding the repeater rate, can be the classical communication time to acknowledge the success of entanglement distillation \cite{Jiang2009}. To overcome this bottleneck, the quantum repeater using quantum error-correcting codes \cite{Jiang2009,Munro2010} was developed. In these protocols, classical communication is only needed between the neighboring repeater stations.

In this paper we investigate the difference in the secret key rates between the quantum repeater using distillation (\textit{generic quantum repeater}) and the quantum repeater using quantum error correcting codes (\textit{encoded quantum repeater}) by employing the analysis developed in \cite{Bratzik2013} for the generic quantum repeater. As a representative for the encoded quantum repeater we choose \cite{Jiang2009}. We analyze the trade-off between the communication time and the needed resources. For this purpose, we modify the analysis given in \cite{Jiang2009} by using a concatenated error model for which we obtain a bound for the fidelity. Our model does not need a fault-tolerant preparation of the initial states, as it handles input pairs which are depolarized states with fidelity $F_0<1$. This approach saves resources, compared to \cite{Jiang2009}, where multi-qubit errors are either suppressed or avoided via distillation. Furthermore, we show how to correct additional errors in the encoded connection procedure and we develop a decoding algorithm suitable for quantum key distribution. We derive the rate for generating entangled pairs with the encoded quantum repeater and use it to calculate secret key rates.

The paper is organized as follows: in Sec.~\ref{sec:QEC} we first briefly review the repeater scheme from Ref.~\cite{Jiang2009}. Furthermore, the error model and its effect on the quantum states is introduced. We show which errors can be corrected during the encoded connection step and develop an appropriate decoding procedure. We then derive the repeater rate for the encoded quantum repeater scheme, i.e., the average number of entangled Bell pairs per second. In Sec.~\ref{sec:SKR}, we review the generic quantum repeater, which uses distillation instead of quantum error-correcting codes. Then we provide the threshold parameters for the encoded quantum repeater for obtaining a nonzero secret key rate. We continue to calculate the optimal secret key rate of these quantum repeater protocols and point out where our analysis differs from the original proposal of the encoded quantum repeater scheme \cite{Jiang2009}.
We then present a short analysis of the cost function which was introduced recently in \cite{Muralidharan2013}. We conclude in Sec.~\ref{sec:concl}.

\section{\label{sec:QEC}Encoded quantum repeater scheme and secret key rates}
In this section we introduce for the quantum repeater using encoding \cite{Jiang2009} (in the following  called \textit{encoded quantum repeater}) a generic error model, show its effect in the different steps of this quantum repeater model and derive the repeater rate, which is needed to calculate the secret key rate. 

\subsection{Principles of the encoded QR}
\begin{figure}[ht!]
\includegraphics[width=0.8\columnwidth]{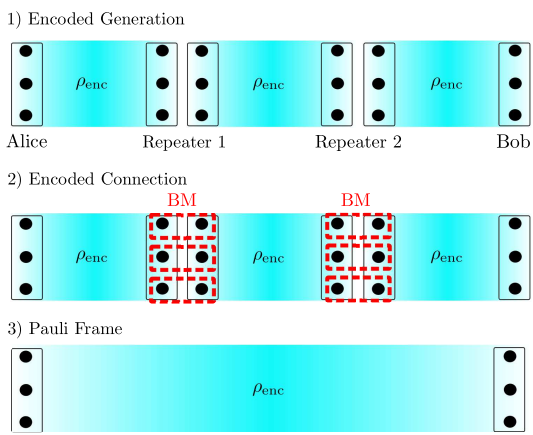}
\caption{\label{fig:encQR_setup}(Color online) Setup of the encoded quantum repeater (adapted from \cite{Jiang2009}), BM stands for Bell measurement. In the ideal case $\rho_{\rm enc}=\proj{\tilde{\phi}^+}$ with $\ket{\tilde{\phi}^+}$ defined in Eq.~\eqref{eq:encBell2}.}
\end{figure}

The principle of the encoded quantum repeater is depicted in Fig.~\ref{fig:encQR_setup}: the first step is to distribute Bell pairs between the neighboring repeater stations, it follows the encoding operations [step 1) in Fig.~\ref{fig:encQR_setup}] and an entanglement swapping scheme [step 2)] which allows to obtain error information. The error information reveals the necessary rotation in order to get a specific encoded Bell pair in the end [step 3)]. The details for entanglement swapping and classical error correction are described in \cite{Jiang2009}. The encoded quantum repeater was developed for any CSS-code \cite{Calderbank1996,Steane1996}.

For simplicity, we will consider the three-qubit repetition code throughout this paper\footnote{We will see in Sec.~\ref{subsec:cost} that this simple code leads to a good ratio of the secret key rate and the required resources.}. In the ideal case the encoded state $\rho_{\rm enc}$, shared between the repeater stations $R_i$ and $R_{i+1}$ at step 1) is of the form $\rho_{\rm enc}=\proj{\tilde{\phi}^+}$ with
\begin{equation}
\ket{\tilde{\phi}^+}=\frac{1}{\sqrt{2}}\ket{000}_{R_i}\ket{000}_{R_{i+1}}+\ket{111}_{R_i}\ket{111}_{R_{i+1}}.
\label{eq:encBell2}
\end{equation}

\subsubsection{\label{sec:error}Error models}
Analogously to \cite{Dur1999,Jiang2009}, we will employ the depolarizing error model for all two-qubit gates, thus the unitary operation $U_{i,j}$ acting on qubits $i$ and $j$ is replaced according to the following map $\Lambda(\rho)$:
\begin{equation}
U_{i,j}^{}\rho U_{i,j}^\dagger\rightarrow(1-\beta)U_{i,j}^{}\rho U_{i,j}^\dagger+\frac{\beta}{4}\trz{\rho}{i,j} \otimes \mathds{1}_{i,j}=:\Lambda(\rho),
\label{eq:cnoterr}
\end{equation}
where $\beta$ is the gate error parameter. We further assume no misalignment and errorfree one-qubit operations\footnote{This assumption about errorfree single-qubit rotations can be made as these rotations can be implemented in the classical postprocessing (application of bit-flips on the measurement data).}. We define
\begin{equation}
U^a:=\id_{\{1,\dots,N\}\setminus\{i_{a},j_{a}\}}\otimes U_{i_{a},j_{a}}, 
\end{equation}
to be the unitary operation acting on the two qubits $(i_{a},j_{a})$ and the identity on the remaining $N-2$ qubits. The vector $\vec{U}=\{U^1,...,U^n\}$ defines the sequence of applications of the unitary operations: first one applies the gate $U^1$ on the qubits $i_1,j_1$, then $U^2$ and so on. For our analytical analysis we will approximate the concatenation of $n$ two-qubit gates by assuming that not more than one gate acts in a faulty way (which corresponds to an expansion in $\beta$, keeping only terms in zeroth and first order). Normalization is guaranteed by adding the worst case density matrix (i.e., the identity) for the remaining probability.

Thus, the resulting map $\Lambda_{\rm conc}(\rho)$ is:
\begin{eqnarray}
\Lambda_{\rm conc}(\rho)&:=&(1-\beta)^n \left(\prod_{a=1}^n U^a\right)\rho \left(\prod_{a=1}^nU^a\right)^\dagger+ n\beta(1-\beta)^{n-1}\tilde{\rho}\nonumber\\
&+&p \frac{\id_{d}}{d},
\label{eq:state1}
\end{eqnarray}
where $d=\dim(\rho)$, $\id_{d}$ is the $d\times d$-identity matrix,  and $p=1-(1-\beta)^n-n\beta (1-\beta)^{n-1}$. For small $\beta$ we can expand $p$ in $p\approx\frac{n(n-1)}{2}\beta^2-n\binom{n-1}{2}\beta^3$, thus $p$ is in the order of $\beta^2$ for an appropriate $n$. The normalized state $\tilde{\rho}$ is given by the map $\Lambda_{\rm 1-faulty}(\rho)$
\begin{eqnarray}
\tilde{\rho}&=&\frac{1}{n}\sum_{a=1}^{n}\left(\prod_{b=a+1}^{n}U^b\right)f\left[(i_a,j_a),\rho,\prod_{c=1}^{a-1} U^c\right]\left(\prod_{b=a+1}^{n}U^b\right)^\dagger\nonumber\\
&=:&\Lambda_{\rm 1-faulty}(\rho),
\label{eq:rhotilde}
\end{eqnarray}
with $f[(i,j), \rho, A]:=\trz{A\rho A^\dagger}{i,j}\otimes \frac{\id_{i,j}}{4}$. Thus, $\tilde{\rho}$ represents the convex combination of states where one gate is replaced by the identity matrix in the corresponding subspace. Instead of the first-order approximation map in Eq.~\eqref{eq:state1} one could use the simpler map
\begin{equation}\tilde{\Lambda}(\rho):=(1-\beta)^n\left(\prod_{a=1}^nU^a\right)\rho \left(\prod_{a=1}^nU^a\right)^\dagger+[1-(1-\beta)^n]\frac{\id_d}{d}.
\label{eq:simple}
\end{equation}
But our analysis shows a distinct improvement in the secret key rate using the map in Eq.~\eqref{eq:state1}.

\subsubsection{Encoded state generation}
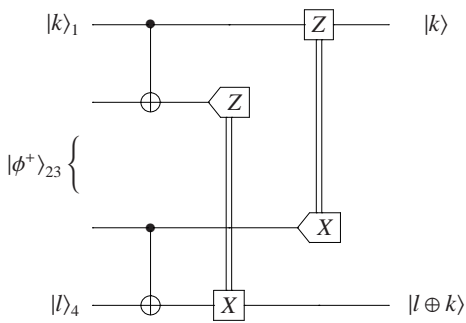
\begin{figure}[ht]
\[
\Qcircuit @C=2em @R=2em {
\lstick{\ket{k}_1} & \ctrl{1} & \qw&\gate{Z}\cwx[3]&\qw& \ket{k} \\
& \targ & \measuretab{Z}\cwx[3] \\
\lstick{\ket{\phi^+}_{23}\Bigg\{}& & \\
& \ctrl{1} & \qw &\measuretab{X}\\
\lstick{\ket{l}_4}&\targ &\gate{X}&\qw&\qw&\ket{l\oplus k}
}
\] 
\caption{\label{fig:tCNOT}Teleportation-based \cnot, see \cite{Jiang2007b}.}
\end{figure}

\begin{figure}[ht!]
\includegraphics[width=\columnwidth]{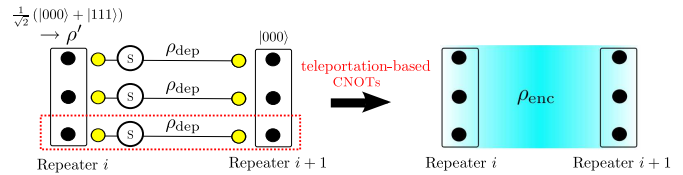}
\caption{\label{fig:encQR}(Color online) Generation of the encoded state $\rho_{enc}$ (see text).}
\end{figure}

The first step performed in the encoded quantum repeater is to generate the encoded Bell state of Eq.~\eqref{eq:encBell2} between all repeater stations. Thus, the encoded Bell state is denoted as (we drop the indices $R_i$ and $R_{i+1}$ for better readability) 
\begin{equation}
\ket{\tilde{\phi}^+}=\frac{1}{\sqrt{2}}\ket{000000}+\ket{111111}.
\label{eq:encBell}
\end{equation}
To generate this state one starts with the state $\frac{1}{\sqrt{2}}\left(\ket{000}+\ket{111}\right)$ at one repeater station and with $\ket{000}$ at the other and applies a \textit{teleportation-based controlled-\textsc{not}} (\textsc{cnot}) \cite{Gottesman1999,Zhou2000,Jiang2007b}:
\begin{eqnarray}
&&\frac{1}{\sqrt{2}}\left(\ket{000}+\ket{111}\right)\otimes\ket{000}\nonumber\\
&&\underrightarrow{\textsc{cnot}}\frac{1}{\sqrt{2}}\ket{000000}+\ket{111111}.
\end{eqnarray}
The teleportation-based \cnot~consists of multiple gates and requires Bell pairs as a resources, as shown in Fig.~\ref{fig:tCNOT}.
The setup in the encoded quantum repeater is shown in Fig.~\ref{fig:encQR}. Between the repeater stations we have a source ($S$) of Bell states. The teleportation-based \cnot~is marked by the red dashed box, i.e., qubits $1$ and $2$ in Fig.~\ref{fig:tCNOT} correspond to one black and one yellow (light grey) qubit of repeater station $i$ and qubits $3$ and $4$ to the qubits of repeater station $i+1$. In total we have three teleportation based \cnot s (see Fig.~\ref{fig:encQR}). The total circuit in Fig.~\ref{fig:encQR} consists of 6 two-qubit gates, thus we apply the concatenation of gates as described in the map of Eq.~\eqref{eq:state1}. The distributed Bell states are depolarized due to imperfections in the source resulting in $\rho_{\rm dep}$ with fidelity $F_0$:
\begin{equation}
\rho_{\rm dep}=F_0\proj{\phi^+}+\frac{1-F_0}{3}\left(\id_{4}-\proj{\phi^+}\right).
\label{eq:rhoDep}
\end{equation}
Different to the proposal in \cite{Jiang2009}, we assume that the initial states do not have fidelity $F_0$ almost one.
Due to the depolarizing error of the quantum operations, see Eq.~\eqref{eq:cnoterr}, the state $\frac{1}{\sqrt{2}}\left(\ket{000}+\ket{111}\right)$ (repeater station $i$ in Fig.~\ref{fig:encQR}) transforms to\footnote{We obtain this state by applying two faulty \cnot s ($\cnota{1}{3}$ and $\cnota{1}{2}$) on the state $\frac{1}{\sqrt{2}}\left(\ket{0}+\ket{1}\right)\ket{00}$. 
} 

\begin{eqnarray}
\rho'&=&\frac{1}{2}\left[1+\beta\left(\frac{\beta}{2}-\frac{5}{4}\right)\right]\left(\Pi_{\ket{000}}+\Pi_{\ket{111}}\right)\nonumber\\
&&+\frac{1}{2}\left(1-\beta\right)^2\left(\ketbra{000}{111}+\ketbra{111}{000}\right)\nonumber\\
&&+\frac{\beta}{4}\left(\frac{3}{2}-\beta\right)\left(\Pi_{\ket{101}}+\Pi_{\ket{010}}\right)\nonumber\\
&&+\frac{\beta}{8}\left(\Pi_{\ket{001}}+\Pi_{\ket{110}}+\Pi_{\ket{100}}+\Pi_{\ket{011}}\right),
\label{eq:rhoPrime}
\end{eqnarray}
where $\Pi_{\ket{klm}}=\proj{klm}$. The state $\rho_{\rm enc}$ after all operations is lengthy and will not be given explicitly here.

\subsubsection{\label{sec:EncC}Encoded connection}
\begin{figure}[ht]
 \[
\Qcircuit @C=2em @R=2em {
 & \ctrl{1} &\measuretab{X}\\
& \targ &\measuretab{Z} 
}
\]
\caption{\label{fig:BM}Circuit for a Bell measurement.} 
\end{figure}
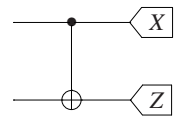

In the second step we perform three pairwise Bell measurements (BM in Fig.~\ref{fig:encQR_setup} and Fig.~\ref{fig:BM}) in the repeater station in order to connect two encoded Bell pairs.
The results of the Bell measurements determine the encoded Bell state (see \cite{Jiang2009} for further explanation). As the three-qubit repetition code is used, we can correct up to one bit-flip error in the measurement results. The bit-flip error is corrected via \textit{majority voting}. Note that the error is corrected classically in the measurement results; no quantum operation is performed on the state.

We denote the map for the total encoded connection $\mathcal{C}:\mathcal{B}(\mathcal{H})\rightarrow \mathcal{B}(\mathcal{H}')$, where $\mathcal{B}(\mathcal{H})$ is the space of bounded operators with $\dim{\mathcal{H}}=2^{12}$ and $\dim{\mathcal{H}'}=2^6$. The map consists of the following procedures: Bell measurement, correction of the measurement results and application of the corresponding Pauli matrices in order to obtain $\ket{\tilde{\phi}^+}$ in the end. Note that the action of the Pauli matrices can be replaced by applying bit-flips on the measurement data of the final state. In the following we determine all states $\ket{\varphi_i}\in\mathcal{H}$, such that an application of perfect gates would lead to the correct state, i.e.,
\begin{equation}
\mathcal{C}^{\text{perf}}(\proj{\varphi_i})=\proj{\tilde{\phi}^+}.
\end{equation}

\paragraph{Correctable errors for ideal \cnot}
The three-qubit repetition code can correct single bit-flip errors. However, due to the properties of the Bell measurement given in Fig.~\ref{fig:BM} one can correct more errors, which were not considered in the analysis in \cite{Jiang2009}.

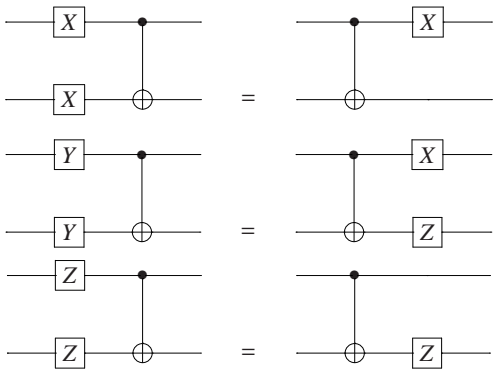
\begin{figure}[ht!]
 \[
\Qcircuit @C=2em @R=2em {
&\gate{X} & \ctrl{1} & \qw&&& \ctrl{1}& \gate{X} &\qw\\
&\gate{X}& \targ & \qw &=&& \targ &\qw&\qw
}
\]
 \[
\Qcircuit @C=2em @R=2em {
&\gate{Y} & \ctrl{1} & \qw&&& \ctrl{1}& \gate{X} &\qw\\
&\gate{Y}& \targ & \qw &=&& \targ &\gate{Z}&\qw
}
\]
 \[
\Qcircuit @C=2em @R=2em {
&\gate{Z} & \ctrl{1} & \qw&&& \ctrl{1}& \qw &\qw\\
&\gate{Z}& \targ & \qw &=&& \targ &\gate{Z}&\qw
}
\]
\caption{\label{fig:gates}Commutation rules for the \cnot~gates, see \cite{Nielsen2000}.}
\end{figure}

If an X error on the control and the target qubit occurs before a perfect \cnot~gate, the resulting error after the application of the \cnot~is an X-error on the control qubit (see, e.g., \cite{Nielsen2000}). As the Bell measurement is performed by applying a measurement in the $X$-basis ($Z$-basis) on the control (target) qubit, this error does not corrupt the measurement result. The same holds also for $Y$- and $Z$-errors if they appear both in the control and target qubit of one \cnot~ (see Fig.~\ref{fig:gates}).

\begin{figure}[ht!]
\includegraphics[width=0.6\columnwidth,angle=270]{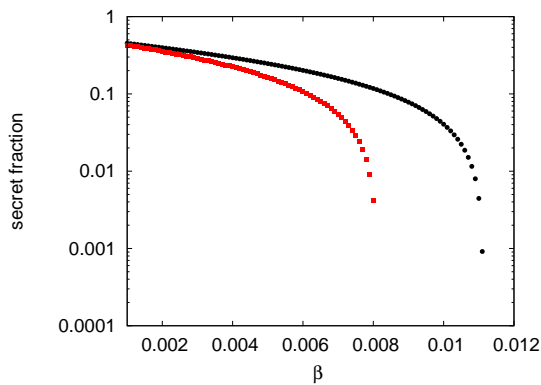}
\caption{\label{fig:SKR_98_r=1} The secret fraction [Eq.~\eqref{eq:sfrac}] plotted as a function of the gate error $\beta$ including all errors according to Eq.~\eqref{eq:errors} (black circles) and those where only single bit-flip errors were correctable (red squares), $F_0=0.98$, one repeater station ($r=1$). }
\end{figure}

Our analysis shows that considering the correlated \mbox{$X$-,} \mbox{$Y$-,} and $Z$-errors leads to a substantially higher error tolerance for obtaining a nonzero secret key rate. As the error in the gates is only important for the secret fraction in Eq.~\eqref{eq:sfrac}, the secret fraction for different initial fidelities as a function of the gate error parameter $\beta$ is displayed in Fig.~\ref{fig:SKR_98_r=1}. We find that including the correlated error makes the secret key rate more noise-tolerant, where the amount of improvement depends on the initial fidelity $F_0$.

We denote the control (target) qubits of the $i$-th \cnot~gate as $c_i$ ($t_i$), thus we can correct errors from the set
\begin{equation}
E=\{X_{c_i}X_{t_i},Y_{c_i}Y_{t_i},Z_{c_i}Z_{t_i},\id_{c_i}\id_{t_i},\id_{c_i}X_{t_i},X_{c_i}\id_{t_i}\}.
\label{eq:errors}
\end{equation} 
Each of the six pairs from the set of errors can happen at one of the three \cnot~gates, thus we have $6^3=216$ combinations. But some of these combinations have to be excluded, e.g., cases like  $\id_{c_1}X_{t_1}\id_{c_2}X_{t_2}\id_{c_3}\id_{t_3}$ are not allowed, as they would lead to a wrong measurement outcome for the majority voting (two $X$-errors cannot be corrected by the three-qubit repetition code). Excluding these cases $160$ combinations remain. We count the number of permutations for three \cnot~gates: Fixing, e.g., the combination $X_{c_i}X_{t_i}Y_{c_j}Y_{t_j}Z_{c_k}Z_{t_k}$ with $i\neq j\neq k\in\{1,2,3\}$ one has 6 possible permutations of $\{1,2,3\}$. In total we have $160\times6=960$ possible combinations of correctable errors. Let us denote these correctable error as $E^i$, with $i=1,\dots,960$, and the set that contains them as $E_{\rm corr}$.

The probability of successful entanglement swappings is defined by the overlap of the states to be swapped ($\rho_{\rm enc}$, see Fig.~\ref{fig:encQR_setup}) with the correctable states $\ket{\varphi_i}$. These states $\ket{\varphi_i}$ are computed by the action of correctable errors $E^i$ from the set $E_{\rm corr}$ onto the states $\ket{\tilde{\phi}^+}\otimes\ket{\tilde{\phi}^+}$:
\begin{equation}
\ket{\varphi_i}:=E^i\left(\ket{\tilde{\phi}^+}\otimes\ket{\tilde{\phi}^+}\right),\; i=1,\dots,960.
\label{eq:phi}
\end{equation}
Out of the $960$ correctable states we have $64$ distinct orthogonal states, denoted as $\ket{\tilde{\varphi_i}}$. The probability of successful entanglement swapping is thus:
\begin{equation}
p_s=\sum_{i=1}^{64} \braccket{\tilde{\varphi_i}}{\rho_{\rm enc}\otimes \rho_{\rm enc}}{\tilde{\varphi_i}},
\end{equation}
where $\rho_{\rm enc}$ is the output state after the teleportation-based \cnot.
This probability holds for one repeater station. For $r$ repeaters we can bound the success probability by\footnote{Note that this estimate is a lower bound, as with more entanglement swappings we could certainly correct more errors.}
\begin{equation}
P_r=(p_s)^r,
\label{eq:Fr}
\end{equation}
because we assumed that the errors can be corrected independently. The final state after entanglement swapping is given by:
\begin{equation}
\rho_{\rm swap}^{\rm ideal}(r)=P_r \proj{\tilde{\phi^+}}+\frac{(1-P_r)}{2^6-1}\left(\id_{2^6}-\proj{\tilde{\phi^+}}\right),
\label{stateES}
\end{equation}
i.e.,
\begin{equation}
\mathcal{C}^{\text{perf}}\left(\rho_{\rm enc}\otimes\rho_{\rm enc}\right)=\rho_{\rm swap}^{\rm ideal}(r). 
\end{equation}
The state given in Eq.~\eqref{stateES} is an estimate; with probability $P_r$ we obtain the perfect state and with probability $1-P_r$ we obtain the completely mixed state without the perfect state.

\paragraph{Nonideal \cnot}
The nonideal \cnot~operation in the Bell measurements is obtained by using the noise model of Eq.~\eqref{eq:state1}:
\begin{eqnarray}
\Lambda_{\textsc{cnot}}(\rho)&=&(1-\beta)^3U\rho U^\dagger+3\beta(1-\beta)^2\tilde{\rho}\nonumber\\
&&+\left[1-\left(1-\beta\right)^3-3\beta(1-\beta)^2\right]\frac{\id_{2^6}}{2^6},
\label{eq:BMMap}
\end{eqnarray}
where $U$ is the concatenation of ideal \cnot s.
The state $\tilde{\rho}$, see Eq.~\eqref{eq:rhotilde}, is the convex combination of states where one of the three \cnot~gates is replaced by the identity matrix in the corresponding subspace [see Eq.~\eqref{eq:cnoterr}]. The map for encoded connection with imperfect \cnot s acting on the correctable states $\ket{\tilde{\varphi_i}}$ with $i=1,\dots,64$ is
\begin{eqnarray}
\mathcal{C}^{\text{imperf}}\left(\proj{\varphi_i}\right)&=&(1-\beta)^3\proj{\tilde{\phi}^+}\nonumber\\
&+& 3\beta(1-\beta)^2\frac{1}{2}\left(\Pi_{\ket{000000}}+\Pi_{\ket{111111}}\right)\nonumber\\
&+&\left[1-\left(1-\beta\right)^3-3\beta(1-\beta)^2\right]\frac{\id_{2^6}}{2^6}\nonumber\\
&=:&\rho_{s}.
\end{eqnarray}
The resulting state for imperfect \cnot s after one round of entanglement swapping is given by:
\begin{eqnarray}
&&\mathcal{C}^{\text{imperf}}\left(\rho_{\rm enc}\otimes\rho_{\rm enc}\right)\nonumber\\
&=&P_1\rho_s+\frac{1-P_1}{2^6-1}\left(\id_{2^6}-\proj{\tilde{\phi}^+}\right).
\end{eqnarray}

For more than one repeater station, we use the approximation 
\begin{eqnarray}
\mathcal{C}^{\text{imperf}}\left(\proj{\varphi_i}^{\otimes r}\right)&=&(1-\beta)^{3r}\proj{\tilde{\phi}^+}\nonumber\\
&+& 3^{r}\beta^r(1-\beta)^{2r}\frac{1}{2}\left(\Pi_{\ket{000000}}+\Pi_{\ket{111111}}\right)\nonumber\\
&+&\left[1-\left(1-\beta\right)^{3r}-3^r\beta^r(1-\beta)^{2r}\right]\frac{\id_{2^6}}{2^6}\nonumber\\
&=:&\rho_{s}(r),
\label{eq:rhos}
\end{eqnarray}
which leads to a lower bound on the secret key rate; higher order terms are represented as identity, more useful states could be present.

Finally, the state with $r$ repeater stations after swapping is given by:
\begin{eqnarray}
\rho_{\rm swap}^{\rm nonideal}(r)&=&P_r\rho_s(r)\nonumber\\
&&+\frac{1-P_r}{2^6-1}\left(\id_{2^6}-\proj{\tilde{\phi}^+}\right),
\label{eq:stateDec}
\end{eqnarray}
with $P_r$ given in Eq.~\eqref{eq:Fr}.

\subsubsection{\label{subsec:dec}Decoding and final state}
In order to do quantum key distribution, one needs to decode the state $\rho_{\rm swap}^{\rm nonideal}(r)$ in Eq.~\eqref{eq:stateDec}.

We assume that the decoding procedure is the reverse process of the encoding procedure  (see Fig.~\ref{fig:dec}):
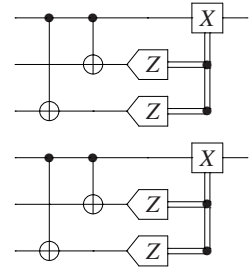
\begin{figure}[ht!]
\[
\Qcircuit @C=1em @R=0.7em {
&\ctrl{2}&\ctrl{1}&\qw&\gate{X}&\qw\\
&\qw&\targ&\measuretab{Z}&\control \cw \cwx\\
&\targ&\qw&\measuretab{Z}&\control \cw \cwx\\
&\ctrl{2}&\ctrl{1}&\qw&\gate{X}&\qw\\
&\qw&\targ&\measuretab{Z}&\control \cw \cwx\\
&\targ&\qw&\measuretab{Z}&\control \cw \cwx
}
\]
\caption{\label{fig:dec}Decoding procedure: The upper three qubits are on Alice's and the lower are on Bob's side.}
\end{figure}
 Alice and Bob each measure two of their three qubits. As we employ the three-qubit repetition code, one can only correct one bit-flip error for Alice and Bob. Other errors cannot be corrected by this code. Depending on their measurement results, Alice and Bob have to correct their qubit with a bit-flip operation. If Alice's and Bob's input state was only subjected to one-qubit bit-flip errors, they only have to correct their qubit if their measurement outcome was $"11"$, which is the case when there was a bit-flip error on the first qubit. If an error occurs in the second or the third qubit, the error does not propagate to the first qubit, thus no correction has to be performed. 

To obtain the final state which is used for quantum key distribution, we take the state after swapping in Eq.~\eqref{eq:stateDec} and perform the decoding operation $\mathcal{D}$:
\begin{equation}
\rho_{\rm dec}=\mathcal{D}\left[\rho_{\rm swap}^{\rm nonideal}(r)\right].
\end{equation}
The decoding map has the following properties:
\begin{subequations}
\label{eqs:prop}
\begin{eqnarray}
&\mathcal{D}\left(\proj{\tilde{\phi}^+}\right)=\proj{\phi^+},\label{eq:prop1}\\
&\mathcal{D}\left(\Pi_{\ket{000000}}+\Pi_{\ket{111111}}\right)=\proj{00}+\proj{11},\label{eq:prop2}\\
&\mathcal{D}\left(\frac{\id_{2^6}}{2^6}\right)
=\frac{1}{4}\proj{\phi^+}+\frac{3}{4}\left[\frac{1}{3}\left(\id_{2^2}-\proj{\phi^+}\right)\right]=\frac{\id_{2^2}}{4}.\label{eq:prop3}
\end{eqnarray}
\end{subequations}
The first property [Eq.~\eqref{eq:prop1}] follows from the action of the \cnot~gates, see Fig.~\ref{fig:dec}:
\begin{eqnarray}
 \ket{\tilde{\phi}^+}\rightarrow \ket{\phi}^+_{1,4}\otimes \ket{0000}_{2,3,5,6},
\end{eqnarray}
where the index denotes the number of qubits.
The second property [Eq.~\eqref{eq:prop2}] can be shown in an analogous way:
\begin{eqnarray}
&&\Pi_{\ket{000000}}+\Pi_{\ket{111111}}\nonumber\\
&\rightarrow& \left(\proj{00}_{1,4}+\proj{11}_{1,4}\right)\otimes \proj{0000}_{2,3,5,6}.
\end{eqnarray}
The last equality [Eq.~\eqref{eq:prop3}] can be verified by inserting the completely mixed state into the decoding map. Using Eqs.~\eqref{eq:rhos} and~\eqref{eq:stateDec} the state after perfect decoding is given by:
\begin{eqnarray}
\rho_{\rm dec}&=&\left[P_r\left(1-\beta\right)^{3r}-\frac{1-P_r}{2^6-1}\right]\proj{\phi^+}\nonumber\\
&&+P_r3^r\beta^r(1-\beta)^{2r}\frac{1}{2}\left(\proj{00}+\proj{11}\right)\nonumber\\
&&+\left[P_r q_r+\frac{(1-P_r)2^6}{2^6-1}\right]\frac{\id_{2^2}}{4},
\label{eq:sfrac}
\end{eqnarray}
with $q_r=1-\left(1-\beta\right)^{3r}-3^r\beta^r(1-\beta)^{2r}$.

We can also include gate errors in the decoding, by again using the error model in Eq.~\eqref{eq:state1}, leading to the final state:
\begin{eqnarray}
\rho_{\rm final}&=&(1-\beta)^4 \rho_{\rm dec}\nonumber\\
&&+4\beta(1-\beta)^3 \rho_{\rm dec}^{\rm nonideal}\nonumber\\
&&+(1-(1-\beta)^4-4\beta(1-\beta)^3)\frac{\id_4}{4},
\label{eq:rhofinal}
\label{eq:rhofinal2}
\end{eqnarray}
where $\rho_{\rm dec}^{\rm nonideal}=\mathcal{D}^{\rm imperf}(\rho_{\rm swap}^{\rm nonideal}(r))$ which is composed of $\Lambda_{\rm 1-faulty}(\rho_{\rm swap}^{\rm nonideal}(r))$, see Eq.~\eqref{eq:rhotilde}, and applying the correcting operation (see Fig.~\ref{fig:dec}) such that
\begin{eqnarray}
&&\mathcal{D}^{\rm imperf}\left(\proj{\tilde{\phi}^+}\right)=\mathcal{D}^{\rm imperf}\left(\frac{1}{2}\left(\Pi_{\ket{000000}}+\Pi_{\ket{111111}}\right)\right)\nonumber\\
&=&\frac{1}{2}\left\{ \left[\frac{3}{8}\left(\proj{00}+\proj{11}\right)+\frac{1}{8}\left(\proj{01}+\proj{10}\right)\right]\right.\nonumber\\
 &&\left.+\frac{\id_4}{4}\nonumber\right\}\\
&=:&\tilde{\rho}'.
\end{eqnarray}
The state $\rho_{\rm dec}^{\rm nonideal}$ is given by
\begin{eqnarray}
\rho_{\rm dec}^{\rm nonideal}&=&P_r\left\{ \left[(1-\beta)^{3r}+3^r\beta^r(1-\beta)^{2r}\right]\tilde{\rho}'\right.\nonumber\\
&&\left.+\{1-[(1-\beta)^{3r}+3^r\beta^r(1-\beta)^{2r}]\}\frac{\id_4}{4} \right\}\nonumber\\
&&+\frac{1-P_r}{2^6-1}\left(2^6\frac{\id_4}{4}-\tilde{\rho}'\right).
\end{eqnarray}
We now use the final state $\rho_{\rm final}$ in Eq.~\eqref{eq:rhofinal} to calculate secret key rates.

\subsection{\label{sec:Rrep}Secret key rate}
Analogously to \cite{Bratzik2013}, we define the secret key rate per memory per second for the repeater to be 
\begin{equation}
K^\nu=R^\nu \frac{r_\infty}{M^\nu},
\label{eq:Ksec}
\end{equation}
where $R^\nu$ is the repeater rate, i.e., the average number of generated entangled Bell pairs per second for the repeater scheme\footnote{By the repeater scheme $\nu$, we mean the scheme mentioned in the introduction: either the encoded quantum repeater ($\nu=\text{QEC}$) or the generic quantum repeater ($\nu=\text{QR}$).} $\nu$, $r_\infty$ is the secret fraction, i.e., the ratio of secret bits and measured bits in the asymptotic limit (Devetak-Winter bound \cite{Devetak2005}) and $M^\nu$ is the number of memories used for each protocol. For the three-qubit repetition code employed here, the number of memories per half node of a repeater station is given by 
\begin{equation}
M^{\rm QEC}=6, 
\label{eq:M}
\end{equation}
as we need six qubits on each side to perform the teleportation-based \cnot~(see Fig.~\ref{fig:encQR}). 
The formula for the secret fraction using the six-state protocol \cite{Bruss1998,Bechmann-Pasquinucci1999} can be found, e.g., in Ref.~\cite{Scarani2009}:
\begin{eqnarray}
\label{eq:6S}
r_\infty&=&1-e_{Z}h\left(\frac{1+(e_{X}-e_{Y})/e_{Z}}{2}\right)\nonumber\\
&&-(1-e_{Z})h\left(\frac{1-(e_X+e_Y+e_Z)/2}{1-e_Z}\right)\nonumber\\
&&-h(e_{Z}),
\label{eq:rsix}
\end{eqnarray}
where the binary Shannon entropy is given by
\begin{equation}
h(p)=-p \log_2 p-(1-p) \log_2 (1-p),
\end{equation} 
and $e_X$, $e_Y$ and $e_Z$  are the error rates in the $X$-, $Y$-, and $Z$-basis, respectively. The analytic form of the error rates for Bell-diagonal states can be found in \cite{Scarani2009}. It is possible to perform the analysis in an analogous way for other QKD-protocols such as the BB84-protocol \cite{Bennett1984}.

The remaining term in the secret key rate is the repeater rate $R^\nu$, which is the average number of generated entangled Bell pairs per second. For its derivation, we first estimate the average waiting time to distribute the Bell pairs needed for the teleportation-based \cnot~(see Fig.~\ref{fig:tCNOT}). The probability of successful generation of one Bell pair over the distance $L_0$ is 
\begin{equation}
P_0=10^{-\alpha L_0/10},
\label{eq:P0}
\end{equation}
with $\alpha=0.17$ dB/km a realistic photon absorption coefficient for telecom fibers. The probability $P_0$ corresponds to the transmittivity of photons in an optical fiber with an attenuation length of $L_{\rm att}=25.5$ km. In \cite{Bernardes2011} the average waiting time for generating $N$ Bell pairs with probability $P_0$ is given by
\begin{equation}
\langle T\rangle_N=T_0 Z_N(P_0),
\end{equation}
where $T_0=L_0/c$ is the fundamental time (where $c=2\times10^5$ km/s is the speed of light in the fiber) and  
\begin{equation}
Z_N(P_0):=\sum_{j=1}^{N}\binom{N}{j}\frac{(-1)^{j+1}}{1-(1-P_0)^j}
\label{eq:ZN}
\end{equation}
is the average number of attempts to connect $N$ pairs, see \cite{Bernardes2011}. For the encoded quantum repeater with $r$ repeater stations we need to establish $3(r+1)$ Bell pairs, thus the repeater rate (the reciprocal value of the average waiting time) for the encoded quantum repeater is given by
\begin{equation}
R^{\rm QEC}=\frac{1}{\langle T \rangle}=\frac{1}{2T_0 Z_{3 (r+1)}(P_0)},
\label{eq:RQEC}
\end{equation}
under the assumption that the entanglement swapping process is deterministic, i.e., the probability of successful entanglement swapping is one.
The factor $2$ in front of the fundamental time $T_0$ accounts for the time needed to send a photon and acknowledge its arrival. Note that no further classical communication is needed in this repeater protocol. This is contrary to the generic quantum repeater, where after distillation and entanglement swapping in each step the success has to be communicated \cite{Bratzik2013}. We will compare this protocol with respect to the secret key rate to the generic quantum repeater which needs much more classical communication.

\section{\label{sec:SKR}Results and comparison of the secret key rates}

In this section, we first investigate the minimally required parameters for obtaining a nonzero secret key for the encoded quantum repeater. Then we find the optimal secret key rate by optimizing over the encoded quantum repeater and the generic quantum repeater model (see \cite{Bratzik2013}).

\subsection{\label{subsec:minipara}Minimal parameters}
Similar to our analysis in \cite{Abruzzo2013}, we derive for the encoded quantum repeater the minimally required parameters, i.e., the initial fidelity $F_0$ [for the depolarized state given in Eq.~\eqref{eq:rhoDep}] and the gate quality $p_G=1-\beta$ [see Eq.~\eqref{eq:state1}], to obtain a nonzero secret key, see Eq.~\eqref{eq:Ksec}. Table~\ref{Tab} summarizes the results of our investigations about the minimal fidelities and gate qualities, which are needed to achieve a nonzero secret key rate.

\begin{table}[ht]
\begin{tabular}{*{1}{*{4}{l}}}
\hline\hline
$r$& $N$& $p_{G,\min}$ & $F_{0,\min}$\\
\hline
$1$ & $1$ & $0.984$ & $0.944$ \\
$3$ & $2$ & $0.993$ & $0.972$ \\
$7$ & $3$ & $0.994$ & $0.981$ \\
$15$ & $4$ & $0.996$ & $0.986$ \\
$31$ & $5$ & $0.997$ & $0.989$ \\
$63$ & $6$ & $0.997$ & $0.991$ \\
$127$ & $7$ & $0.998$ & $0.992$ \\
\hline\hline
\end{tabular}

\caption{\label{Tab}Minimal gate quality $p_{G,\min}$ [see Eq.~\eqref{eq:state1} with $p_G=1-\beta$ and $F_0=1$] and minimal fidelity $F_{0,\min}$ with $p_G=1$ for extracting a secret key for the six-state protocol, see Eq.~\eqref{eq:rsix}, with $r=2^N-1$ repeater stations, where $N$ is the nesting level for entanglement swapping.}
\end{table}

\begin{figure}[ht]
\includegraphics[width=0.6\columnwidth,angle=270]{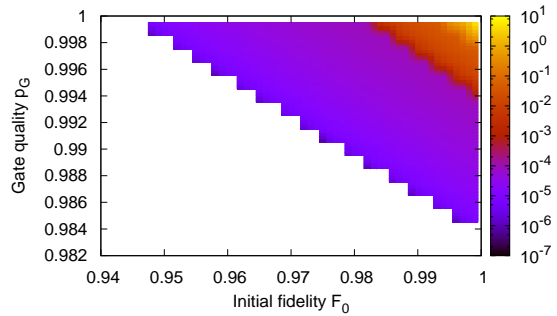}
\caption{\label{fig:SKR_encQR}(Color online) The optimal secret key rate per memory per second, see Eq.~\eqref{eq:Ksec}, as a function of the initial fidelity $F_0$ and the gate quality $p_G$, optimized over the number of repeater stations for the encoded quantum repeater ($L=600$ km). In the white region it is not possible to extract a nonzero secret key.}
\end{figure}

Comparing these numbers to the results for the generic quantum repeater in \cite{Abruzzo2013}, we find that the encoded quantum repeater is less tolerant against gate errors. This can be expected as many gates are needed for generating the encoded Bell pair. Regarding the initial fidelity $F_0$, the encoded quantum repeater also requires fairly good initial Bell states. In Fig.~\ref{fig:SKR_encQR}, we show the secret key rate for the encoded quantum repeater, optimized with respect to the number of repeater stations, as a function of the gate quality $p_G$ and initial fidelity $F_0$, for a fixed distance. We find that a nonzero secret key rate for the encoded quantum repeater is restricted to gate errors below $\beta=0.0165$ and fidelities above $F_0=0.943$.

\subsection{Optimal secret key rates}
The purpose of this paper is to investigate whether the optimal secret key rate is reached for the encoded or the generic quantum repeater. By the generic quantum repeater, we mean the repeater scheme using distillation (either the \textit{Deutsch et al.}~\cite{Deutsch1996} or the \textit{D\"{u}r et al.} \cite{Dur1999} distillation protocols, see \cite{Bratzik2013}). 

\begin{figure}[ht]
\includegraphics[width=0.6\columnwidth,angle=270]{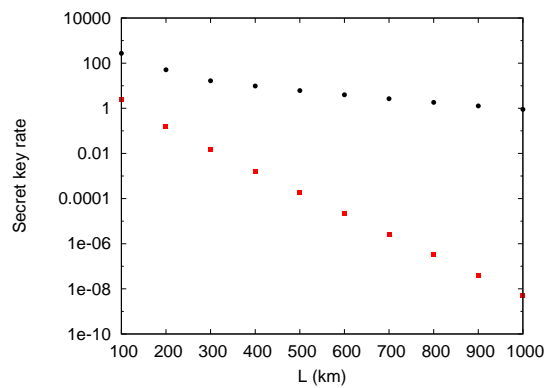}
\caption{\label{fig:SKR_vs_L}(Color online) The optimal secret key rate per memory per second in Eq.~\eqref{eq:Ksec} for the encoded (red squares) and the generic quantum repeater (black circles) as a function of the total distance $L$ (Parameters: $F_0=0.98$, $p_G=0.992$).}
\end{figure}

Figure~\ref{fig:SKR_vs_L} shows the optimal secret key rate per memory per second [Eq.~\eqref{eq:Ksec} for the two different repeater schemes plotted as the function of the distance for some realistic parameters (initial fidelity $F_0=0.98$ and gate quality $p_G=0.992$). We find that the generic quantum repeater leads to an optimal secret key rate that is one order of magnitude better than the encoded quantum repeater. We know from \cite{Bratzik2013} that in this range of parameters it is optimal for the generic quantum repeater to not distill, thus the number of used memories is one in this case. For the encoded quantum repeater, however, we constantly use six memories [see Eq.~\eqref{eq:M}] which reduces the secret key rate by this factor. In the case of no distillation and deterministic entanglement swapping the encoded quantum repeater is not an advantage for the chosen parameters. Also in regimes where distillation is optimal for the generic quantum repeater (see \cite{Bratzik2013}), we find that the encoded repeater is never better.

\subsection{Comparison to the scheme in Jiang et al.}
As mentioned in the introduction, Ref.~\cite{Jiang2009} analyzes the errors of the encoded quantum repeater. Different to their analysis, we allow the initial Bell pairs to have a fidelity $F_0<1$, whereas in the original reference fault-tolerant distillation is assumed which would require additional qubits and operations. Also we do not perform fault-tolerant initialization of the codes which has the following advantages:
our scheme saves resources and no additional measurements have to be realized. Our error model, see Eq.~\eqref{eq:state1}, is a very good estimate, especially for the $\beta$ resulting from our investigations of the minimal parameters ($\beta=10^{-2}-10^{-3}$, see Sec.~\ref{subsec:minipara}): the probability $p$ for the identity is on the order of $10^{-3}$. It means that the distance to a map, where the exact output state is used, is small. We verified this statement for the decoding map (Sec.~\ref{subsec:dec}) and found that the Uhlmann fidelity of the state in Eq.~\eqref{eq:rhofinal2} compared to a state calculated with a map, where all gates with errors are used, is one for the range of $\beta$ given above. - So far, we do not consider memory and measurement errors, but they could be easily implemented in our analysis.

Another difference in the analysis is the repeater rate. In \cite{Jiang2009} it was assumed that the generation rate for undistilled Bell pairs with $m$ qubits\footnote{Note that $m=M/2$ with $M$ given in Eq.~\eqref{eq:M}.} available at each station is given by\footnote{The probability of successful transmission $P_0$ [see Eq.~\eqref{eq:P0}] is equivalent to the expression $\exp(-L_0/L_{\rm att})$ given in \cite{Jiang2009}.}
\begin{equation}
R=m \frac{P_0}{L_0},
\label{eq:RJiang}
\end{equation}
using a perfect overall efficiency for collecting and detecting single photons. This is only an appropriate estimate in the case of infinitely many memories. We considered that exactly $m$ Bell pairs to start the encoding process are needed and one has $m$ memories instead of infinitely many memories available. This results in the estimate with the average waiting time as described by Eq.~\eqref{eq:ZN} (Sec.~\ref{sec:Rrep}) and leads to a decrease of the repeater rate [Eq.~\eqref{eq:RQEC}] by several orders of magnitude\footnote{The repeater rate and thus the secret key rate can be increased by using multiplexing as shown in \cite{Abruzzo2013a}.} compared to Eq.~\eqref{eq:RJiang}.

\subsection{\label{subsec:cost}Cost function}
Recently, in \cite{Muralidharan2013} a quantum repeater scheme was investigated that only uses one-way classical communication without the necessity to herald the successful generation of entangled Bell pairs between the repeater stations.

In \cite{Muralidharan2013}, the cost function was defined to be the minimum number of total memory qubits per secret bit:
\begin{equation}
C=\min_{\nu,N}\frac{2^{N+1}}{K^\nu},
\label{eq:cost}
\end{equation}
where $K$ is the secret key rate as defined in Eq.~\eqref{eq:Ksec}, $N$ is the nesting level and $\nu$ is the index for the chosen protocol. The factor $2^{N+1}$ accounts for the total number of memories (with the number of memories $M$ per half node of the repeater station, see Eq.~\eqref{eq:M}, being implicitly contained in the secret key rate K): $(2^N-1) 2$ is the sum of all memory qubits in the repeater stations ($r=2^N-1$). Adding $2$ memory qubits from the two communicating parties (Alice and Bob) results in $2^{N+1}$ memory qubits in total.  

\begin{figure}[ht]
\includegraphics[width=0.6\columnwidth,angle=270]{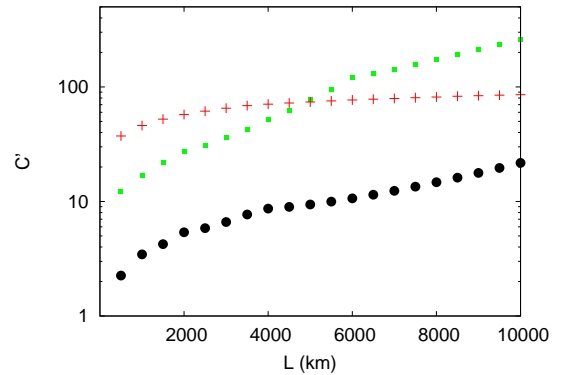}
\caption{\label{fig:cost2_Jiang} (Color online) The cost coefficient ($C'=C/L$) [Eq.~\eqref{eq:cost}] for the encoded (green squares), the generic quantum repeater (black circles) and the quantum repeater protocol presented in \cite{Muralidharan2013} (red crosses, with the effective qubit error $\varepsilon=10^{-4}$, for an explanation see \cite{Muralidharan2013}) as a function of the total distance $L$ (Parameters: $F_0=0.99995$, $p_G=0.9999$, and $T_0=1$ as in \cite{Muralidharan2013}).}
\end{figure}

Figure~\ref{fig:cost2_Jiang} shows the cost coefficient $C'$, which is the cost function $C$ [Eq.~\eqref{eq:cost}] divided by the total length $L$, using the encoded quantum repeater with the three-qubit repetition code, the generic quantum repeater and the repeater protocol presented in \cite{Muralidharan2013}. The optimal distillation protocol for the generic quantum repeater is the \textit{Deutsch et al.}~protocol. We find that up to 5000 km both the generic quantum repeater and the encoded quantum repeater are below the cost coefficient given in \cite{Muralidharan2013}. The generic quantum repeater has a better resource efficiency for all distances. The generic quantum repeater needs less resources and the overhead in classical communication is compensated by fewer numbers of qubit memories used.

\begin{figure}[ht]
\includegraphics[width=0.6\columnwidth,angle=270]{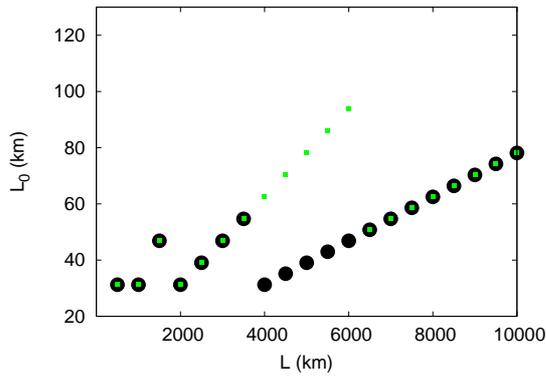}
\caption{\label{fig:cost2_Jiang_L0} (Color online) The optimal distance between the repeater stations $L_0$ for the cost coefficient ($C'=C/L$), see Eq.~\eqref{eq:cost}, for the encoded (green squares) and the generic quantum repeater (black circles) (parameters $F_0=0.99995$ and $p_G=0.9999$, $T_0=1$ as in \cite{Muralidharan2013}) as a function of the total distance $L$.}
\end{figure}

In Fig.~\ref{fig:cost2_Jiang_L0}, we show the optimal distance $L_0$ between the repeater stations for the encoded and the generic quantum repeater as a function of $L$. We point out that $L_0$ is in the order of $30-120$ km, depending on $L$, while the optimal distance in \cite{Muralidharan2013} was given by $1-2$ km. In the generic quantum repeater scheme the total number of required repeater stations is circa $1-2$ orders of magnitude smaller than in \cite{Muralidharan2013}.

\section{\label{sec:concl}Conclusions}
We investigated secret key rates for the so-called \textit{encoded quantum repeater} that utilizes quantum error-correcting codes instead of entanglement distillation as used in the \textit{generic quantum repeater}. The advantage of the former repeater scheme is that the classical communication is minimal as it is needed only for the entanglement generation. Before starting to calculate the secret key rate, we improved the error model for the encoded quantum repeater \cite{Jiang2009} and replaced it by a concatenated one, which leads to a very good estimate of the fidelity. This model permits us to start with Bell pairs with fidelity smaller than one. We also accounted for additional correctable errors for the encoded connection of the encoded Bell states and finally developed a decoding algorithm suitable for the three-qubit repetition code using an analogous error model. We estimated the minimally required parameters for a nonzero secret key and found that for many repeater stations the requirements for the gate quality and the initial fidelity of the Bell pairs are quite demanding. In order to calculate the secret key rate for the encoded quantum repeater, we derived the rate for generating entangled pairs. The comparison of the secret key rates for the encoded and generic quantum repeater showed that the generic scheme is advantageous for the whole considered range of parameters for the gate quality and the initial fidelity. Finally, we calculated the cost function for the repeater schemes studied here. The cost function determines the required resources divided by the secret key rate. We found that the cost function for the schemes analyzed here is better than the scheme presented in \cite{Muralidharan2013}, for a wide range of parameters. Furthermore, the number of repeater stations is circa $1-2$ orders of magnitude smaller than in the latter scheme.

So far, measurement and memory errors are excluded, but can be implemented easily in our analysis. The secret key rate can be improved by applying multiplexing (see, e.g., \cite{Abruzzo2013a}). As we performed the calculations on density matrices, the quantum error-correcting codes used here for the encoded quantum repeater are limited to a small number of qubits. As an example we treated the three-qubit repetition code. We showed that even with this simple quantum error-correcting code we can have an advantage regarding the cost function over complicated schemes using more resources. We conjecture that utilizing more sophisticated codes (see, e.g., \cite{Jiang2009}) does not lead to an increase of the secret key rate, as more resources are needed especially when the encoding is performed with errors.

\begin{acknowledgments}
The authors thank Michael Epping, Sreraman Muralidharan, and Liang Jiang for discussions. The authors acknowledge financial support by the German Federal Ministry of Education and Research (BMBF, project QuOReP).
\end{acknowledgments}
\bibliography{library}

\begin{thebibliography}{26}%
\makeatletter
\providecommand \@ifxundefined [1]{%
 \@ifx{#1\undefined}
}%
\providecommand \@ifnum [1]{%
 \ifnum #1\expandafter \@firstoftwo
 \else \expandafter \@secondoftwo
 \fi
}%
\providecommand \@ifx [1]{%
 \ifx #1\expandafter \@firstoftwo
 \else \expandafter \@secondoftwo
 \fi
}%
\providecommand \natexlab [1]{#1}%
\providecommand \enquote  [1]{``#1''}%
\providecommand \bibnamefont  [1]{#1}%
\providecommand \bibfnamefont [1]{#1}%
\providecommand \citenamefont [1]{#1}%
\providecommand \href@noop [0]{\@secondoftwo}%
\providecommand \href [0]{\begingroup \@sanitize@url \@href}%
\providecommand \@href[1]{\@@startlink{#1}\@@href}%
\providecommand \@@href[1]{\endgroup#1\@@endlink}%
\providecommand \@sanitize@url [0]{\catcode `\\12\catcode `\$12\catcode
  `\&12\catcode `\#12\catcode `\^12\catcode `\_12\catcode `\%12\relax}%
\providecommand \@@startlink[1]{}%
\providecommand \@@endlink[0]{}%
\providecommand \url  [0]{\begingroup\@sanitize@url \@url }%
\providecommand \@url [1]{\endgroup\@href {#1}{\urlprefix }}%
\providecommand \urlprefix  [0]{URL }%
\providecommand \Eprint [0]{\href }%
\providecommand \doibase [0]{http://dx.doi.org/}%
\providecommand \selectlanguage [0]{\@gobble}%
\providecommand \bibinfo  [0]{\@secondoftwo}%
\providecommand \bibfield  [0]{\@secondoftwo}%
\providecommand \translation [1]{[#1]}%
\providecommand \BibitemOpen [0]{}%
\providecommand \bibitemStop [0]{}%
\providecommand \bibitemNoStop [0]{.\EOS\space}%
\providecommand \EOS [0]{\spacefactor3000\relax}%
\providecommand \BibitemShut  [1]{\csname bibitem#1\endcsname}%
\let\auto@bib@innerbib\@empty
\bibitem [{\citenamefont {Briegel}\ \emph {et~al.}(1998)\citenamefont
  {Briegel}, \citenamefont {D\"{u}r}, \citenamefont {Cirac},\ and\
  \citenamefont {Zoller}}]{Briegel1998}%
  \BibitemOpen
  \bibfield  {author} {\bibinfo {author} {\bibfnamefont {H.-J.}\ \bibnamefont
  {Briegel}}, \bibinfo {author} {\bibfnamefont {W.}~\bibnamefont {D\"{u}r}},
  \bibinfo {author} {\bibfnamefont {J.}~\bibnamefont {Cirac}}, \ and\ \bibinfo
  {author} {\bibfnamefont {P.}~\bibnamefont {Zoller}},\ }\href {\doibase
  10.1103/PhysRevLett.81.5932} {\bibfield  {journal} {\bibinfo  {journal}
  {Physical Review Letters}\ }\textbf {\bibinfo {volume} {81}},\ \bibinfo
  {pages} {5932} (\bibinfo {year} {1998})}\BibitemShut {NoStop}%
\bibitem [{\citenamefont {D\"{u}r}\ \emph {et~al.}(1999)\citenamefont
  {D\"{u}r}, \citenamefont {Briegel}, \citenamefont {Cirac},\ and\
  \citenamefont {Zoller}}]{Dur1999}%
  \BibitemOpen
  \bibfield  {author} {\bibinfo {author} {\bibfnamefont {W.}~\bibnamefont
  {D\"{u}r}}, \bibinfo {author} {\bibfnamefont {H.-J.}\ \bibnamefont
  {Briegel}}, \bibinfo {author} {\bibfnamefont {J.}~\bibnamefont {Cirac}}, \
  and\ \bibinfo {author} {\bibfnamefont {P.}~\bibnamefont {Zoller}},\ }\href
  {\doibase 10.1103/PhysRevA.59.169} {\bibfield  {journal} {\bibinfo  {journal}
  {Physical Review A}\ }\textbf {\bibinfo {volume} {59}},\ \bibinfo {pages}
  {169} (\bibinfo {year} {1999})}\BibitemShut {NoStop}%
\bibitem [{\citenamefont {\.{Z}ukowski}\ \emph {et~al.}(1993)\citenamefont
  {\.{Z}ukowski}, \citenamefont {Zeilinger}, \citenamefont {Horne},\ and\
  \citenamefont {Ekert}}]{Zukowski1993}%
  \BibitemOpen
  \bibfield  {author} {\bibinfo {author} {\bibfnamefont {M.}~\bibnamefont
  {\.{Z}ukowski}}, \bibinfo {author} {\bibfnamefont {A.}~\bibnamefont
  {Zeilinger}}, \bibinfo {author} {\bibfnamefont {M.}~\bibnamefont {Horne}}, \
  and\ \bibinfo {author} {\bibfnamefont {A.}~\bibnamefont {Ekert}},\ }\href
  {\doibase 10.1103/PhysRevLett.71.4287} {\bibfield  {journal} {\bibinfo
  {journal} {Physical Review Letters}\ }\textbf {\bibinfo {volume} {71}},\
  \bibinfo {pages} {4287} (\bibinfo {year} {1993})}\BibitemShut {NoStop}%
\bibitem [{\citenamefont {Bennett}\ \emph {et~al.}(1996)\citenamefont
  {Bennett}, \citenamefont {Brassard}, \citenamefont {Popescu}, \citenamefont
  {Schumacher}, \citenamefont {Smolin},\ and\ \citenamefont
  {Wootters}}]{Bennett1996a}%
  \BibitemOpen
  \bibfield  {author} {\bibinfo {author} {\bibfnamefont {C.~H.}\ \bibnamefont
  {Bennett}}, \bibinfo {author} {\bibfnamefont {G.}~\bibnamefont {Brassard}},
  \bibinfo {author} {\bibfnamefont {S.}~\bibnamefont {Popescu}}, \bibinfo
  {author} {\bibfnamefont {B.}~\bibnamefont {Schumacher}}, \bibinfo {author}
  {\bibfnamefont {J.}~\bibnamefont {Smolin}}, \ and\ \bibinfo {author}
  {\bibfnamefont {W.}~\bibnamefont {Wootters}},\ }\href {\doibase
  10.1103/PhysRevLett.76.722} {\bibfield  {journal} {\bibinfo  {journal}
  {Physical Review Letters}\ }\textbf {\bibinfo {volume} {76}},\ \bibinfo
  {pages} {722} (\bibinfo {year} {1996})}\BibitemShut {NoStop}%
\bibitem [{\citenamefont {Deutsch}\ \emph {et~al.}(1996)\citenamefont
  {Deutsch}, \citenamefont {Ekert}, \citenamefont {Jozsa}, \citenamefont
  {Macchiavello}, \citenamefont {Popescu},\ and\ \citenamefont
  {Sanpera}}]{Deutsch1996}%
  \BibitemOpen
  \bibfield  {author} {\bibinfo {author} {\bibfnamefont {D.}~\bibnamefont
  {Deutsch}}, \bibinfo {author} {\bibfnamefont {A.~K.}\ \bibnamefont {Ekert}},
  \bibinfo {author} {\bibfnamefont {R.}~\bibnamefont {Jozsa}}, \bibinfo
  {author} {\bibfnamefont {C.}~\bibnamefont {Macchiavello}}, \bibinfo {author}
  {\bibfnamefont {S.}~\bibnamefont {Popescu}}, \ and\ \bibinfo {author}
  {\bibfnamefont {A.}~\bibnamefont {Sanpera}},\ }\href {\doibase
  10.1103/PhysRevLett.77.2818} {\bibfield  {journal} {\bibinfo  {journal}
  {Physical Review Letters}\ }\textbf {\bibinfo {volume} {77}},\ \bibinfo
  {pages} {2818} (\bibinfo {year} {1996})}\BibitemShut {NoStop}%
\bibitem [{\citenamefont {Duan}\ \emph {et~al.}(2001)\citenamefont {Duan},
  \citenamefont {Lukin}, \citenamefont {Cirac},\ and\ \citenamefont
  {Zoller}}]{Duan2001}%
  \BibitemOpen
  \bibfield  {author} {\bibinfo {author} {\bibfnamefont {L.~M.}\ \bibnamefont
  {Duan}}, \bibinfo {author} {\bibfnamefont {M.~D.}\ \bibnamefont {Lukin}},
  \bibinfo {author} {\bibfnamefont {J.~I.}\ \bibnamefont {Cirac}}, \ and\
  \bibinfo {author} {\bibfnamefont {P.}~\bibnamefont {Zoller}},\ }\href
  {http://dx.doi.org/10.1038/35106500} {\bibfield  {journal} {\bibinfo
  {journal} {Nature}\ }\textbf {\bibinfo {volume} {414}},\ \bibinfo {pages}
  {413} (\bibinfo {year} {2001})}\BibitemShut {NoStop}%
\bibitem [{\citenamefont {van Loock}\ \emph {et~al.}(2006)\citenamefont {van
  Loock}, \citenamefont {Ladd}, \citenamefont {Sanaka}, \citenamefont
  {Yamaguchi}, \citenamefont {Nemoto}, \citenamefont {Munro},\ and\
  \citenamefont {Yamamoto}}]{VanLoock2006}%
  \BibitemOpen
  \bibfield  {author} {\bibinfo {author} {\bibfnamefont {P.}~\bibnamefont {van
  Loock}}, \bibinfo {author} {\bibfnamefont {T.}~\bibnamefont {Ladd}}, \bibinfo
  {author} {\bibfnamefont {K.}~\bibnamefont {Sanaka}}, \bibinfo {author}
  {\bibfnamefont {F.}~\bibnamefont {Yamaguchi}}, \bibinfo {author}
  {\bibfnamefont {K.}~\bibnamefont {Nemoto}}, \bibinfo {author} {\bibfnamefont
  {W.~J.}\ \bibnamefont {Munro}}, \ and\ \bibinfo {author} {\bibfnamefont
  {Y.}~\bibnamefont {Yamamoto}},\ }\href {\doibase
  10.1103/PhysRevLett.96.240501} {\bibfield  {journal} {\bibinfo  {journal}
  {Physical Review Letters}\ }\textbf {\bibinfo {volume} {96}},\ \bibinfo
  {pages} {240501} (\bibinfo {year} {2006})}\BibitemShut {NoStop}%
\bibitem [{\citenamefont {Sangouard}\ \emph {et~al.}(2011)\citenamefont
  {Sangouard}, \citenamefont {Simon}, \citenamefont {de~Riedmatten},\ and\
  \citenamefont {Gisin}}]{Sangouard2011}%
  \BibitemOpen
  \bibfield  {author} {\bibinfo {author} {\bibfnamefont {N.}~\bibnamefont
  {Sangouard}}, \bibinfo {author} {\bibfnamefont {C.}~\bibnamefont {Simon}},
  \bibinfo {author} {\bibfnamefont {H.}~\bibnamefont {de~Riedmatten}}, \ and\
  \bibinfo {author} {\bibfnamefont {N.}~\bibnamefont {Gisin}},\ }\href
  {\doibase 10.1103/RevModPhys.83.33} {\bibfield  {journal} {\bibinfo
  {journal} {Review of Modern Physics}\ }\textbf {\bibinfo {volume} {83}},\
  \bibinfo {pages} {33} (\bibinfo {year} {2011})}\BibitemShut {NoStop}%
\bibitem [{\citenamefont {Abruzzo}\ \emph
  {et~al.}(2013{\natexlab{a}})\citenamefont {Abruzzo}, \citenamefont {Bratzik},
  \citenamefont {Bernardes}, \citenamefont {Kampermann}, \citenamefont {van
  Loock},\ and\ \citenamefont {Bru\ss}}]{Abruzzo2013}%
  \BibitemOpen
  \bibfield  {author} {\bibinfo {author} {\bibfnamefont {S.}~\bibnamefont
  {Abruzzo}}, \bibinfo {author} {\bibfnamefont {S.}~\bibnamefont {Bratzik}},
  \bibinfo {author} {\bibfnamefont {N.~K.}\ \bibnamefont {Bernardes}}, \bibinfo
  {author} {\bibfnamefont {H.}~\bibnamefont {Kampermann}}, \bibinfo {author}
  {\bibfnamefont {P.}~\bibnamefont {van Loock}}, \ and\ \bibinfo {author}
  {\bibfnamefont {D.}~\bibnamefont {Bru\ss}},\ }\href {\doibase
  10.1103/PhysRevA.87.052315} {\bibfield  {journal} {\bibinfo  {journal}
  {Physical Review A}\ }\textbf {\bibinfo {volume} {87}},\ \bibinfo {pages}
  {052315} (\bibinfo {year} {2013}{\natexlab{a}})}\BibitemShut {NoStop}%
\bibitem [{\citenamefont {Bratzik}\ \emph {et~al.}(2013)\citenamefont
  {Bratzik}, \citenamefont {Abruzzo}, \citenamefont {Kampermann},\ and\
  \citenamefont {Bru\ss}}]{Bratzik2013}%
  \BibitemOpen
  \bibfield  {author} {\bibinfo {author} {\bibfnamefont {S.}~\bibnamefont
  {Bratzik}}, \bibinfo {author} {\bibfnamefont {S.}~\bibnamefont {Abruzzo}},
  \bibinfo {author} {\bibfnamefont {H.}~\bibnamefont {Kampermann}}, \ and\
  \bibinfo {author} {\bibfnamefont {D.}~\bibnamefont {Bru\ss}},\ }\href
  {\doibase 10.1103/PhysRevA.87.062335} {\bibfield  {journal} {\bibinfo
  {journal} {Physical Review A}\ }\textbf {\bibinfo {volume} {87}},\ \bibinfo
  {pages} {062335} (\bibinfo {year} {2013})}\BibitemShut {NoStop}%
\bibitem [{\citenamefont {Jiang}\ \emph {et~al.}(2009)\citenamefont {Jiang},
  \citenamefont {Taylor}, \citenamefont {Nemoto}, \citenamefont {Munro},
  \citenamefont {{Van Meter}},\ and\ \citenamefont {Lukin}}]{Jiang2009}%
  \BibitemOpen
  \bibfield  {author} {\bibinfo {author} {\bibfnamefont {L.}~\bibnamefont
  {Jiang}}, \bibinfo {author} {\bibfnamefont {J.}~\bibnamefont {Taylor}},
  \bibinfo {author} {\bibfnamefont {K.}~\bibnamefont {Nemoto}}, \bibinfo
  {author} {\bibfnamefont {W.~J.}\ \bibnamefont {Munro}}, \bibinfo {author}
  {\bibfnamefont {R.}~\bibnamefont {{Van Meter}}}, \ and\ \bibinfo {author}
  {\bibfnamefont {M.~D.}\ \bibnamefont {Lukin}},\ }\href {\doibase
  10.1103/PhysRevA.79.032325} {\bibfield  {journal} {\bibinfo  {journal}
  {Physical Review A}\ }\textbf {\bibinfo {volume} {79}},\ \bibinfo {pages}
  {032325} (\bibinfo {year} {2009})}\BibitemShut {NoStop}%
\bibitem [{\citenamefont {Munro}\ \emph {et~al.}(2010)\citenamefont {Munro},
  \citenamefont {Harrison}, \citenamefont {Stephens}, \citenamefont {Devitt},\
  and\ \citenamefont {Nemoto}}]{Munro2010}%
  \BibitemOpen
  \bibfield  {author} {\bibinfo {author} {\bibfnamefont {W.~J.}\ \bibnamefont
  {Munro}}, \bibinfo {author} {\bibfnamefont {K.~A.}\ \bibnamefont {Harrison}},
  \bibinfo {author} {\bibfnamefont {A.~M.}\ \bibnamefont {Stephens}}, \bibinfo
  {author} {\bibfnamefont {S.~J.}\ \bibnamefont {Devitt}}, \ and\ \bibinfo
  {author} {\bibfnamefont {K.}~\bibnamefont {Nemoto}},\ }\href {\doibase
  10.1038/nphoton.2010.213} {\bibfield  {journal} {\bibinfo  {journal} {Nature
  Photonics}\ }\textbf {\bibinfo {volume} {4}},\ \bibinfo {pages} {792}
  (\bibinfo {year} {2010})}\BibitemShut {NoStop}%
\bibitem [{\citenamefont {Muralidharan}\ \emph {et~al.}(2013)\citenamefont
  {Muralidharan}, \citenamefont {Kim}, \citenamefont {L\"{u}tkenhaus},
  \citenamefont {Lukin},\ and\ \citenamefont {Jiang}}]{Muralidharan2013}%
  \BibitemOpen
  \bibfield  {author} {\bibinfo {author} {\bibfnamefont {S.}~\bibnamefont
  {Muralidharan}}, \bibinfo {author} {\bibfnamefont {J.}~\bibnamefont {Kim}},
  \bibinfo {author} {\bibfnamefont {N.}~\bibnamefont {L\"{u}tkenhaus}},
  \bibinfo {author} {\bibfnamefont {M.~D.}\ \bibnamefont {Lukin}}, \ and\
  \bibinfo {author} {\bibfnamefont {L.}~\bibnamefont {Jiang}},\ }\href
  {http://arxiv.org/abs/1310.5291} {\  (\bibinfo {year} {2013})},\ \Eprint
  {http://arxiv.org/abs/1310.5291v1} {arXiv:1310.5291v1} \BibitemShut {NoStop}%
\bibitem [{\citenamefont {Calderbank}\ and\ \citenamefont
  {Shor}(1996)}]{Calderbank1996}%
  \BibitemOpen
  \bibfield  {author} {\bibinfo {author} {\bibfnamefont {A.}~\bibnamefont
  {Calderbank}}\ and\ \bibinfo {author} {\bibfnamefont {P.}~\bibnamefont
  {Shor}},\ }\href {\doibase 10.1103/PhysRevA.54.1098} {\bibfield  {journal}
  {\bibinfo  {journal} {Physical Review A}\ }\textbf {\bibinfo {volume} {54}},\
  \bibinfo {pages} {1098} (\bibinfo {year} {1996})}\BibitemShut {NoStop}%
\bibitem [{\citenamefont {Steane}(1996)}]{Steane1996}%
  \BibitemOpen
  \bibfield  {author} {\bibinfo {author} {\bibfnamefont {A.}~\bibnamefont
  {Steane}},\ }\href {\doibase 10.1098/rspa.1996.0136} {\bibfield  {journal}
  {\bibinfo  {journal} {Proc. R. Soc. A}\ }\textbf {\bibinfo {volume} {452}},\
  \bibinfo {pages} {2551} (\bibinfo {year} {1996})}\BibitemShut {NoStop}%
\bibitem [{\citenamefont {Gottesman}\ and\ \citenamefont
  {Chuang}(1999)}]{Gottesman1999}%
  \BibitemOpen
  \bibfield  {author} {\bibinfo {author} {\bibfnamefont {D.}~\bibnamefont
  {Gottesman}}\ and\ \bibinfo {author} {\bibfnamefont {I.}~\bibnamefont
  {Chuang}},\ }\href {\doibase 10.1038/46503} {\bibfield  {journal} {\bibinfo
  {journal} {Nature}\ }\textbf {\bibinfo {volume} {402}},\ \bibinfo {pages}
  {390} (\bibinfo {year} {1999})}\BibitemShut {NoStop}%
\bibitem [{\citenamefont {Zhou}\ \emph {et~al.}(2000)\citenamefont {Zhou},
  \citenamefont {Leung},\ and\ \citenamefont {Chuang}}]{Zhou2000}%
  \BibitemOpen
  \bibfield  {author} {\bibinfo {author} {\bibfnamefont {X.}~\bibnamefont
  {Zhou}}, \bibinfo {author} {\bibfnamefont {D.}~\bibnamefont {Leung}}, \ and\
  \bibinfo {author} {\bibfnamefont {I.}~\bibnamefont {Chuang}},\ }\href
  {\doibase 10.1103/PhysRevA.62.052316} {\bibfield  {journal} {\bibinfo
  {journal} {Physical Review A}\ }\textbf {\bibinfo {volume} {62}},\ \bibinfo
  {pages} {052316} (\bibinfo {year} {2000})}\BibitemShut {NoStop}%
\bibitem [{\citenamefont {Jiang}\ \emph {et~al.}(2007)\citenamefont {Jiang},
  \citenamefont {Taylor}, \citenamefont {S\o{}rensen},\ and\ \citenamefont
  {Lukin}}]{Jiang2007b}%
  \BibitemOpen
  \bibfield  {author} {\bibinfo {author} {\bibfnamefont {L.}~\bibnamefont
  {Jiang}}, \bibinfo {author} {\bibfnamefont {J.}~\bibnamefont {Taylor}},
  \bibinfo {author} {\bibfnamefont {A.}~\bibnamefont {S\o{}rensen}}, \ and\
  \bibinfo {author} {\bibfnamefont {M.~D.}\ \bibnamefont {Lukin}},\ }\href
  {\doibase 10.1103/PhysRevA.76.062323} {\bibfield  {journal} {\bibinfo
  {journal} {Physical Review A}\ }\textbf {\bibinfo {volume} {76}},\ \bibinfo
  {pages} {062323} (\bibinfo {year} {2007})}\BibitemShut {NoStop}%
\bibitem [{\citenamefont {Nielsen}\ and\ \citenamefont
  {Chuang}(2000)}]{Nielsen2000}%
  \BibitemOpen
  \bibfield  {author} {\bibinfo {author} {\bibfnamefont {M.}~\bibnamefont
  {Nielsen}}\ and\ \bibinfo {author} {\bibfnamefont {I.}~\bibnamefont
  {Chuang}},\ }\href@noop {} {\emph {\bibinfo {title} {{Quantum computation and
  quantum information}}}}\ (\bibinfo  {publisher} {Cambridge University
  Press},\ \bibinfo {address} {Cambridge},\ \bibinfo {year} {2000})\BibitemShut
  {NoStop}%
\bibitem [{\citenamefont {Devetak}\ and\ \citenamefont
  {Winter}(2005)}]{Devetak2005}%
  \BibitemOpen
  \bibfield  {author} {\bibinfo {author} {\bibfnamefont {I.}~\bibnamefont
  {Devetak}}\ and\ \bibinfo {author} {\bibfnamefont {A.}~\bibnamefont
  {Winter}},\ }\href {\doibase 10.1098/rspa.2004.1372} {\bibfield  {journal}
  {\bibinfo  {journal} {Proc. R. Soc. A}\ }\textbf {\bibinfo {volume} {461}},\
  \bibinfo {pages} {207} (\bibinfo {year} {2005})}\BibitemShut {NoStop}%
\bibitem [{\citenamefont {Bru\ss}(1998)}]{Bruss1998}%
  \BibitemOpen
  \bibfield  {author} {\bibinfo {author} {\bibfnamefont {D.}~\bibnamefont
  {Bru\ss}},\ }\href {\doibase 10.1103/PhysRevLett.81.3018} {\bibfield
  {journal} {\bibinfo  {journal} {Physical Review Letters}\ }\textbf {\bibinfo
  {volume} {81}},\ \bibinfo {pages} {3018} (\bibinfo {year}
  {1998})}\BibitemShut {NoStop}%
\bibitem [{\citenamefont {Bechmann-Pasquinucci}\ and\ \citenamefont
  {Gisin}(1999)}]{Bechmann-Pasquinucci1999}%
  \BibitemOpen
  \bibfield  {author} {\bibinfo {author} {\bibfnamefont {H.}~\bibnamefont
  {Bechmann-Pasquinucci}}\ and\ \bibinfo {author} {\bibfnamefont
  {N.}~\bibnamefont {Gisin}},\ }\href {\doibase 10.1103/PhysRevA.59.4238}
  {\bibfield  {journal} {\bibinfo  {journal} {Physical Review A}\ }\textbf
  {\bibinfo {volume} {59}},\ \bibinfo {pages} {4238} (\bibinfo {year}
  {1999})}\BibitemShut {NoStop}%
\bibitem [{\citenamefont {Scarani}\ \emph {et~al.}(2009)\citenamefont
  {Scarani}, \citenamefont {Bechmann-Pasquinucci}, \citenamefont {Cerf},
  \citenamefont {Du\v{s}ek}, \citenamefont {L\"{u}tkenhaus},\ and\
  \citenamefont {Peev}}]{Scarani2009}%
  \BibitemOpen
  \bibfield  {author} {\bibinfo {author} {\bibfnamefont {V.}~\bibnamefont
  {Scarani}}, \bibinfo {author} {\bibfnamefont {H.}~\bibnamefont
  {Bechmann-Pasquinucci}}, \bibinfo {author} {\bibfnamefont {N.~J.}\
  \bibnamefont {Cerf}}, \bibinfo {author} {\bibfnamefont {M.}~\bibnamefont
  {Du\v{s}ek}}, \bibinfo {author} {\bibfnamefont {N.}~\bibnamefont
  {L\"{u}tkenhaus}}, \ and\ \bibinfo {author} {\bibfnamefont {M.}~\bibnamefont
  {Peev}},\ }\href {\doibase 10.1103/RevModPhys.81.1301} {\bibfield  {journal}
  {\bibinfo  {journal} {Reviews of Modern Physics}\ }\textbf {\bibinfo {volume}
  {81}},\ \bibinfo {pages} {1301} (\bibinfo {year} {2009})}\BibitemShut
  {NoStop}%
\bibitem [{\citenamefont {Bennett}\ and\ \citenamefont
  {Brassard}(1984)}]{Bennett1984}%
  \BibitemOpen
  \bibfield  {author} {\bibinfo {author} {\bibfnamefont {C.~H.}\ \bibnamefont
  {Bennett}}\ and\ \bibinfo {author} {\bibfnamefont {G.}~\bibnamefont
  {Brassard}},\ }in\ \href@noop {} {\emph {\bibinfo {booktitle} {Proceedings of
  IEEE International Conference on Computers, Systems and Signal Processing}}}\
  (\bibinfo  {publisher} {IEEE, New York},\ \bibinfo {year} {1984})\ p.\
  \bibinfo {pages} {175}\BibitemShut {NoStop}%
\bibitem [{\citenamefont {Bernardes}\ \emph {et~al.}(2011)\citenamefont
  {Bernardes}, \citenamefont {Praxmeyer},\ and\ \citenamefont {van
  Loock}}]{Bernardes2011}%
  \BibitemOpen
  \bibfield  {author} {\bibinfo {author} {\bibfnamefont {N.~K.}\ \bibnamefont
  {Bernardes}}, \bibinfo {author} {\bibfnamefont {L.}~\bibnamefont
  {Praxmeyer}}, \ and\ \bibinfo {author} {\bibfnamefont {P.}~\bibnamefont {van
  Loock}},\ }\href {\doibase 10.1103/PhysRevA.83.012323} {\bibfield  {journal}
  {\bibinfo  {journal} {Physical Review A}\ }\textbf {\bibinfo {volume} {83}},\
  \bibinfo {pages} {012323} (\bibinfo {year} {2011})}\BibitemShut {NoStop}%
\bibitem [{\citenamefont {Abruzzo}\ \emph
  {et~al.}(2013{\natexlab{b}})\citenamefont {Abruzzo}, \citenamefont
  {Kampermann},\ and\ \citenamefont {Bru\ss}}]{Abruzzo2013a}%
  \BibitemOpen
  \bibfield  {author} {\bibinfo {author} {\bibfnamefont {S.}~\bibnamefont
  {Abruzzo}}, \bibinfo {author} {\bibfnamefont {H.}~\bibnamefont {Kampermann}},
  \ and\ \bibinfo {author} {\bibfnamefont {D.}~\bibnamefont {Bru\ss}},\ }\href
  {\doibase 10.1103/PhysRevA.89.012303} {\bibfield  {journal} {\bibinfo
  {journal} {Physical Review A}\ }\textbf {\bibinfo {volume} {89}},\ \bibinfo
  {pages} {012303} (\bibinfo {year} {2013}{\natexlab{b}})}\BibitemShut
  {NoStop}%
\end{thebibliography}%

\end{document}